\documentclass[conference]{IEEEtran}
\IEEEoverridecommandlockouts

\ifCLASSINFOpdf
   \usepackage[pdftex]{graphicx}
   \graphicspath{illustrations/pdf/}
  \DeclareGraphicsExtensions{.pdf}
\else
\fi
\usepackage{balance}

\usepackage{booktabs}
\usepackage{array}
\usepackage{mdwtab}
\usepackage{url}
\usepackage{paralist}
\usepackage{hyperref}

\newcommand{\dist}{-5mm}

\begin{document}
%
% paper title
% can use linebreaks \\ within to get better formatting as desired
\title{Why Feature Dependencies Challenge the Requirements Engineering of
Automotive Systems: An Empirical Study}

% author names and affiliations
% use a multiple column layout for up to two different
% affiliations

\author{
\IEEEauthorblockN{Andreas Vogelsang} 
\IEEEauthorblockA{Institut f\"ur Informatik\\
Technische Universit\"at M\"unchen\\
Boltzmannstr. 3, 85748 Garching, Germany\\
vogelsan@in.tum.de}
\and
\IEEEauthorblockN{Steffen Fuhrmann} 
\IEEEauthorblockA{BMW Group\\
Driving Dynamics\\
Dimensioning Functions Driving Dynamics and Driver Assistance\\
Knorrstr. 147, 80788 M\"unchen, Germany\\
steffen.fuhrmann@bmw.de}

\thanks{This work was partly funded by the German Federal Ministry of
Education and Research (BMBF), grant ``FoMoStA, 01IS12028''.}
}

% make the title area
\maketitle

\begin{abstract}
Functional dependencies and feature interactions in automotive software
systems are a major source of erroneous and deficient behavior. To overcome
these problems, many approaches exist that focus on modeling these functional
dependencies in early stages of system design. 
However, there are only few empirical studies that report on the extent of such
dependencies in industrial software systems and how they are considered in an
industrial development context.
In this paper, we analyze the
functional architecture of a real automotive software system with the aim
to assess the extent, awareness and importance of interactions between features of a
future vehicle.
Our results show that within the functional architecture at least 85\% of the
analyzed vehicle features depend on each other. 
They furthermore show that the developers are not aware of a large number of
these dependencies when they are modeled solely on an architectural level.
Therefore, the developers mention the need for a more precise specification
of feature interactions, e.g., for the execution of comprehensive impact analyses.
These results
challenge the current development methods and emphasize the need for an
extensive modeling of features and their dependencies in requirements
engineering.
\end{abstract}


% For peer review papers, you can put extra information on the cover
% page as needed:
% \ifCLASSOPTIONpeerreview
% \begin{center} \bfseries EDICS Category: 3-BBND \end{center}
% \fi
%
% For peerreview papers, this IEEEtran command inserts a page break and
% creates the second title. It will be ignored for other modes.
\IEEEpeerreviewmaketitle

\section{Introduction}
The behavior of software-intensive embedded systems is characterized by its features and functions.
Many model-based specification techniques for software-intensive systems utilize
this notion of a system feature in order to structure a specification
(e.g.,~\cite{Jackson98,Kang90,Schaetz08}).
Although the notions in the different approaches differ slightly and sometimes synonyms such
as ``system function'', ``feature'', or ``user function'' are used, the
approaches agree on structuring a specification into subparts that contain extracts of the functionality as it is
perceived by the user or any other environmental system. 
We will call these parts \emph{system features} or \emph{vehicle features} in
the remainder of the paper.

Dependencies and interactions between these features are a major challenge in
the industrial development of software-intensive systems~\cite{Zave01}. They increase
the complexity of the system and frequently entail unwanted and deficient system behavior~\cite{Benz10}.
Nevertheless, feature interactions are rarely considered in specifications of
industrial systems.
This leads to increased efforts in late development phases like integration or
system test when these errors typically are revealed~\cite{Benz10}.
We use the terms \emph{functional dependencies} and \emph{feature interaction}
synonymously in this paper.

These dependencies play an important role especially in the development of multifunctional
systems such as automotive software systems~\cite{Broy06,BroyKPS07}). However, dependencies induced by
multifunctionality are a major challenge even for the development of embedded
systems in general. 

In order to handle functional dependencies, many approaches exist, which model
such dependencies in early phases of the development process, i.e., in the specification
or the system design
(e.g.,~\cite{Jackson98,Kang90,Heitmeyer98}).

\subsection{Problem Statement}
Existing approaches towards modeling of functional dependencies have been
validated in specific examples and applications. However, there is little empirical
data on the extent and distribution of such dependencies in industrial software
systems and their consideration in an industrial development context. It is
therefore not possible to thoroughly assess the impact and influence of functional
dependencies and feature interactions on the development of modern software
systems. A comprehensive understanding of the awareness and importance of 
such dependencies in system development is further mandatory to evaluate
existing modeling approaches.

In a previous study concerning the extent and characteristics of functional
dependencies in automotive software systems, we observed a large number and
a complex nature of such dependencies~\cite{Vogelsang12}.
For the further investigation of functional dependencies and the validation of
our earlier results, larger systems with a more networked architecture should be
evaluated. Furthermore, the quantitative analysis of our former study neglects
a qualitative assessment of the feature interactions. A qualitative study is thus necessary
to discover the impact on development methods and processes.

\subsection{Research Objective}
The purpose of our two-phase, sequential mixed methods study is to obtain
quantitative results from a sample and then follow up with a few
individuals to explore those results in depth. In the first phase,
a quantitative research question will address the extent of feature
dependencies in a modern automotive software system. 
%CHANGE
We aim at the identification of value-based dependencies between vehicle features.
We therefore extract data flow dependencies from the functional architecture of
a software system and assess them concerning their value on a vehicle level.
%CHANGE
In the second phase, qualitative interviews will be used to probe the awareness
of the found dependencies and discover the importance for automotive
software developers.

\subsection{Contribution}
We analyzed the functional architecture of a real automotive
software system in order to contribute
\begin{compactitem}
\item data on the amount and distribution of functional dependencies between vehicle features,
\item an evaluation of how dependencies are considered throughout the automotive development process.
\end{compactitem}
The results of our study motivate the use of more extensive modeling techniques
for features and their dependencies.

\subsection{Context}
\label{sec:context}
The study focuses on automotive software systems and was executed at the BMW
Group, a German manufacturer of premium automobiles and motorcycles.
We analyzed the functional architecture
of driving dynamics and driver assistance systems that will be implemented in a future
\emph{sports utility vehicle (SUV)}. The functional architecture consists of
vehicle features, which are grouped into certain feature groups building a
hierarchy of vehicle features. The atomic vehicle features in this hierarchy
are realized by a network of logical components that we refer to as leaf functions.

%CHANGE
In our context, vehicle features and leaf functions build the
central elements for the specification of functional requirements. Subject to their level of detail,
requirements must apply either to the definition of a vehicle feature or a leaf
function. Dependencies within the functional architecture therefore entail dependencies between functional requirements. 
%CHANGE

Within the driving dynamics and driver assistance domain, an in-house developed
database supports the design and implementation of the functional architecture. 
Based on this data backbone, a model-based development approach ensures the
realization of the functional architecture by program code, sensors and actuators.
For the analyses in our study, we used a specific dataset
that we extracted from the described data backbone.

\section{Related Work}
Except from our previous study~\cite{Vogelsang12} and to the best of our knowledge,
there is no comparable work on empirical data or
analyses of realistic automotive or embedded systems with the focus on
dependencies between system features. However, there is a lot of work on
approaches that try to model or specify such dependencies.

Functional dependencies and feature interactions have been
extensively investigated in the telecommunication domain~\cite{Calder00}.
Jackson and Zave~\cite{Jackson98} introduced \emph{Distributed Feature
Composition (DFC)} as a modular, service-oriented architecture for applications
in the telecommunication domain. DFC relies on the notion that a user service
request can be composed of a set of smaller features, which are arranged in a
\emph{pipes-and-filters} architectural style. This architecture is especially
designed for modeling interactions between different features.

Classical approaches like UML use cases, activities or
sequences~\cite{Fowler00} specify system features more or less in isolation.
Dependencies between system features are neglected. This makes it hard to
reason about functionality that arises from the interplay of multiple system
features.

A well-known specification technique for requirements is the
\emph{software cost reduction (SCR)} method~\cite{Heitmeyer98}. In SCR,
requirements are specified by a set of specification tables. The developers of
SCR also noticed that understanding the relationship between
different parts of a specification can be difficult,
especially for large specifications~\cite{HeitmeyerKL97}. Therefore, a Dependency Graph Browser in their tool displays the dependencies
among the variables in a given specification.

\section{Study Design}
In this section, we formulate the research questions, describe
the study object as well as the data
collection and analysis procedures. We conclude with a description of validity
procedures.

\subsection{Research Questions}
The study examines the amount of functional dependencies in automotive software systems and how
dependencies are handled in the development process. We assess the awareness and
importance of functional dependencies to justify the application of feature modeling approaches.
We structured our study with the help of three research questions. 

RQ 1: \textit{What is the overall extent and distribution of dependencies
between vehicle features?}

The relevance of functional dependencies can be motivated inter alia by an analysis
of the overall number of interactions. We define a dependency
between vehicle features as an influence on the behavior of a
vehicle feature by the state or data of another vehicle feature. 

RQ 2: \textit{To what extent are developers aware of functional dependencies?}

Developers of automotive systems are not necessarily aware of existing dependencies
and interactions. We want to identify existing feature dependencies that are unknown
to developers as well as known dependencies that are not represented within the
functional architecture and assess them with regard to their plausibility.

RQ 3: \textit{How important is a comprehensive understanding of functional dependencies
and feature interactions?}

We have to investigate existing feature interactions concerning their importance for
the development process and design decisions in order to reason an extensive
modeling of feature interactions.

\subsection{Study Object}
\label{sec:study_object}
In our study we analyzed an automotive software system of a future 
vehicle and especially its functional architecture. Within the
functional architecture we focused on the driving dynamics and driver
assistance domain. The system comprises 94 vehicle features 
and a total of 325 leaf functions. Leaf functions may be
used for the realization of more than one vehicle feature.

\begin{figure}[!t]
\centering
\includegraphics[width=1\columnwidth]{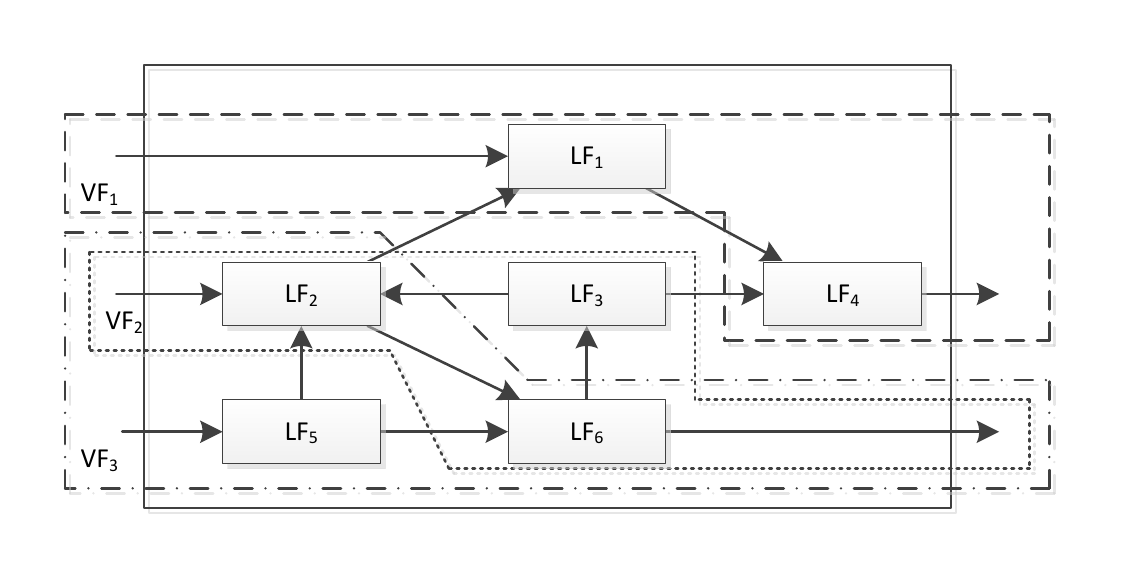}
\caption{The leaf functions (rectangles) are connected by data channels (black arrows) 
and form a functional architecture of the system (outer rectangle). Vehicle
features crosscut this architecture by the set of leaf functions that contribute to their realization (dashed forms).}
\label{fig:functionalarchitecture}
\vspace*{\dist}
\end{figure}

Leaf functions describe the realization/implementation of a
vehicle feature in a purely logical fashion, i.e. without
any information about the hardware the system runs on.
A network of leaf functions describes the steps that
are necessary to transform the input data into the desired
output data. An example for a system that consists of
3 vehicle features, which are realized by a network of 6
leaf functions is illustrated in \figurename~\ref{fig:functionalarchitecture}.
The leaf functions are afterwards assigned to specific software components,
which execute the behavior of the leaf functions. As a final step, these
software components are deployed to a set of electronic computing units.

The relation between a vehicle feature and a leaf function in the context of
this study is the following: A vehicle feature is realized by a set of leaf functions
that are arranged in a data-flow network. A leaf function
can contribute to the realization of a set of vehicle
features. Thus, there is an n~:~m relation between vehicle
features and leaf functions. The set of all leaf functions and their
connections form the functional architecture of the system. The vehicle
features crosscut this architecture by the set of leaf functions that contribute
to their realization (see \figurename~\ref{fig:functionalarchitecture}).

\subsection{Data Collection Procedures}
\begin{figure}[!t]
\centering
\includegraphics[width=0.9\columnwidth]{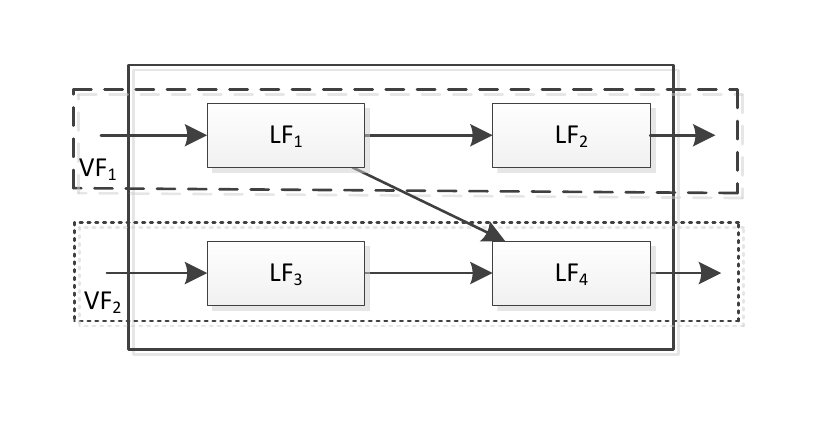}
\caption{The vehicle feature $\mathit{VF_2}$ depends on the vehicle feature $\mathit{VF_1}$,
since the leaf function $\mathit{LF_1}$ (part of $\mathit{VF_1}$) sends values
to the leaf function $\mathit{LF_4}$ (part of $\mathit{VF_2}$). Thus, the
behavior of $\mathit{VF_2}$ depends on data of $\mathit{VF_1}$.}
\label{fig:data_dependency}
\vspace*{\dist}
\end{figure}

\begin{figure}[!t]
\centering
\includegraphics[width=0.9\columnwidth]{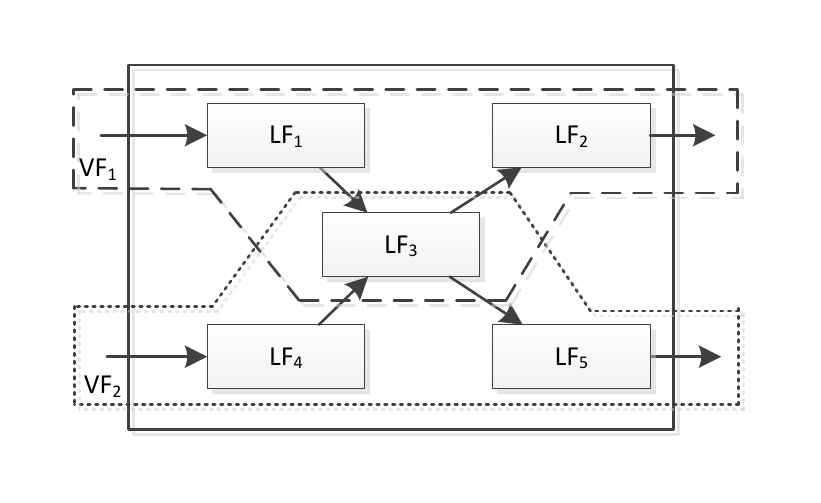}
\caption{Vehicle features $\mathit{VF_1}$ and $\mathit{VF_2}$ share the leaf function $\mathit{LF_3}$. This might indicate a dependency
between the vehicle features. However, this cannot be verified without further knowledge about the
behavior of $\mathit{LF_3}$.}
\label{fig:reuse_dependency}
\vspace*{\dist}
\end{figure}

For the reliable acquisition of data, we need a precise definition of what we consider as a dependency
between vehicle features. Our initial informal definition states that a vehicle
feature $\mathit{VF_1}$ depends on another vehicle feature $\mathit{VF_2}$ if
its behavior is influenced not only by its primary inputs but also by the state
or data of $\mathit{VF_2}$. Therefore, the vehicle features have to communicate with each other.
In our study, we distinguish between two different ways of communication between vehicle features.
\begin{compactenum}
\item A leaf function that is part of one vehicle feature has a communication channel to a leaf function
that is part of another vehicle feature (see \figurename~\ref{fig:data_dependency}).
\item A leaf function is related to two or more vehicle features, i.e., two or
more vehicle features share a leaf function (see \figurename~\ref{fig:reuse_dependency}).
\end{compactenum}

However, a real dependency between two vehicle features cannot be derived
definitely from the shared use of a leaf function. In that case, further knowledge about the concrete
behavior of the leaf function is needed in order to identify the possible dependency. In the example of
\figurename~\ref{fig:reuse_dependency}, the influence of the data transmitted
over the channel $\mathit{LF_1}\rightarrow\mathit{LF_3}$ on the data
transmitted over the channel $\mathit{LF_3}\rightarrow\mathit{LF_5}$ can only
be assessed with further knowledge about the behavior of $\mathit{LF_3}$.

As we had no information about the precise behavior of leaf functions in the
context of our study, we
focused on dependencies of type 1, where a leaf function of one vehicle feature
has a communication channel to a leaf function of another vehicle feature.

\begin{figure}[!t]
\centering
\includegraphics[width=0.5\columnwidth]{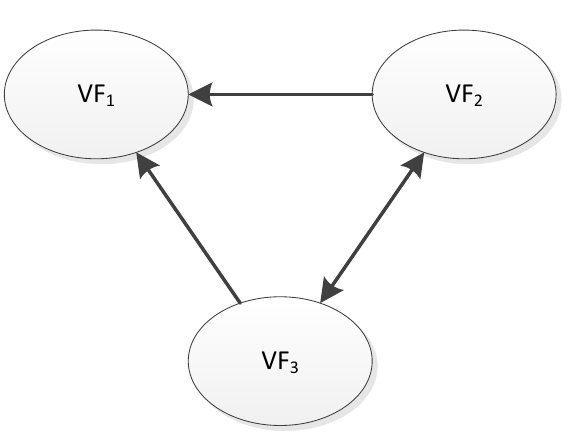}
\caption{The vehicle feature graph extracted from the functional
architecture of \figurename~\ref{fig:functionalarchitecture}.}
\label{fig:vehiclefeatures}
\vspace*{\dist}
\end{figure}

Based on this definition of a dependency between vehicle features, we can
extract a vehicle feature graph from the functional architecture, where each node is a vehicle feature and a
directed edge indicates a dependency between two vehicle features. The
resulting vehicle feature graph for the example of
\figurename~\ref{fig:functionalarchitecture} is illustrated in
\figurename~\ref{fig:vehiclefeatures}. 

%CHANGE
In our study, we extracted the vehicle feature graph by means of a simple tool,
written in Java. The tool parses an exported data set from the company's data
backbone containing a list of features associated with a set of leaf
functions. The tool transforms the data into a graph structure, extracts
the feature dependencies according to the definition given in this section, and
finally outputs a .csv file with the found dependencies. The extraction was
performed fully automated and the complexity of the algorithm is quadratic in
the number of vehicle features and leaf functions. For the observed system, the
extraction took around 3 seconds on a standard laptop.
%CHANGE END

The second part of our study is based on four interviews with feature experts
from the BMW Group, who are involved in the design of the
functional architecture. 

For RQ~2, we confronted the experts with a sample of
feature dependencies from their area of responsibility found by our analysis. We
let the experts classify these dependencies into the following categories:
\begin{compactitem}
  \item \textbf{plausible/implausible:} A dependency is considered as
  \emph{plausible} if the expert finds a functional or physical explanation for this dependency. If the expert has no
  functional or physical explanation for this dependency, it is considered as \emph{implausible}.
  \item \textbf{known/unknown:} A dependency is considered as
  \emph{known} if the expert was aware of this dependency prior to the interview. If the expert was
  not aware of this dependency prior to the interview, it is considered as
  \emph{unknown}.
\end{compactitem}
Overall, we discussed 89 feature dependencies in depth.

For RQ~3, 
% we followed a grounded theory (GT) approach, an explorative research
% method originating from the social sciences, but increasingly popular in
% software engineering research~\cite{Adolph11}. GT is an inductive approach, in which
% interviews are analyzed in order to derive a theory. It aims at discovering new
% perspectives and insights, rather than confirming existing ones. As part of GT,
each interview transcript was analyzed through a process of coding: breaking up
the interviews into smaller coherent units (sentences or paragraphs), and adding
codes (representing key characteristics) to these units. For this purpose, we
asked the experts for experiences in their work where misconceptions
about feature interactions led to errors and increased efforts in the system
design. This part of the interview should provide information about the need for
an extensive modeling of feature interactions.

\subsection{Analysis Procedures}
The procedure for the analysis of the functional architecture varies for the
three research questions:

For RQ~1, we analyzed the vehicle feature graph in order to assess the ratio of
vehicle features that are dependent on another vehicle feature and to count the
number of incoming and outgoing dependencies between vehicle features. We also measured
the dependency fan-in and fan-out as well as the PageRank~\cite{Brin98} for all vehicle
features on the vehicle feature graph in order to see whether dependencies are
distributed equally or if certain vehicle features are more central than others.
Thus, we obtain information about the extent and distribution of functional
dependencies in real automotive software systems.

For RQ~2, we counted the number and ratio of dependencies for each combination
of category values, leading to a 2x2 matrix with the two categories as
dimensions. We especially investigated the ratio of \emph{plausible} feature
dependencies as an indicator for the validity of our quantitative study and the
ratio of \emph{known} feature dependencies as an indicator for the awareness of
feature dependencies in general.

For RQ~3, we developed a coding system with 7 codes structured into 3
categories. We assigned these codes to the units of the interviews. Only codes
that appeared in more than one interview were considered for the study results.
% In the detailed coding schema\footnote{available under:
% \url{http://www4.in.tum.de/~vogelsan/codingschema.pdf}} we give, for each code,
% the name as well as a short one-sentence description and key quotes illustrating the codes.

\subsection{Validity Procedures}
We analyzed the system under investigation at a final stage of the development process where
it was already subject to several architectural reviews and testing procedures.
Therefore, errors and misconceptions in the functional architecture can nearly be ruled out. However, 
our analyses show that vehicle features might differ in the way how they are modeled within
the functional architecture. To ensure validity we presented and discussed the
results with feature experts at the BMW Group, who assessed the found
dependencies concerning their plausibility.
The evaluation of RQ~2 and RQ~3 is based on semi-formal interviews with feature
experts from the BMW Group. In order to get representative results from the
interview partners we selected one expert from each area within the domain of driving dynamics and driver
assistance. These areas are: lateral, longitudinal and vertical dynamics as well as driver
assistance features. The experts were responsible for a number of 12--46
features.

\section{Study Results}

In this section the results of the study are presented.
They are structured according to the defined research
questions.

\subsection{Extent and Distribution of Dependencies (RQ 1)}
Analyzing the vehicle feature graph, we found 1,451 dependencies
between the 94 vehicle features. Only 9 out of the 94 vehicle features were
completely independent from any other vehicle feature. 81 vehicle features were
dependent on another vehicle feature (i.e., had incoming dependencies) and 72
vehicle features had an influence on another vehicle feature (i.e., had
outgoing dependencies).
Table~\ref{tbl:extent} summarizes these results.
There were 234 different logical signals that caused the dependencies.

\begin{table}[!t]
\renewcommand{\arraystretch}{1.3}
\caption{Extent of dependencies in the vehicle feature graph}
\label{tbl:extent}
\centering
%{\scriptsize
\begin{tabular}{|p{5.3cm}||c|c|}\hline
&\bf{Number}&\bf{Ratio}\\ \hline
all VFs & 94 & 100\% \\ \hline
VFs with incoming dependencies & 81 &
86.2\%\\ \hline
VFs with outgoing dependencies & 72 &
76.6\%\\
\hline
VFs with incoming and outgoing dependencies& 68 & 72.3\%\\ \hline
VFs without dependencies& 9 & 9.6\%\\ \hline
\end{tabular}
%}
\end{table}

\begin{figure}[!t]
\centering
\includegraphics[width=0.6\columnwidth]{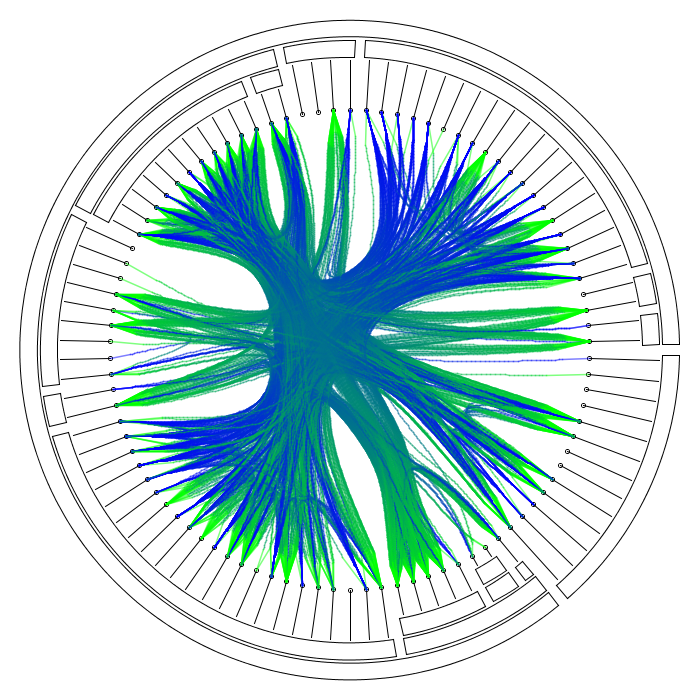}
\caption{Vehicle Features and their dependencies visualized as an Edge Bundle
View. The outer ring represents the hierarchy of vehicle features. Each dot on
the inside of the outer ring is an atomic vehicle feature. The lines indicate a
dependency between two features.}
\label{fig:edgebundleview}
\vspace*{\dist}
\end{figure}

The distribution of the dependencies shows that dependencies between vehicle
features are distributed all over the system. However, there are some vehicle
features that are more central in the sense that they have a large number of
dependencies to other vehicle features. Table~\ref{tbl:distribution} shows that
a vehicle feature depends on up to 48 other vehicle features,
whereas on the other side vehicle features have a maximum of 53 other vehicle
features that they influence, which accounts for 56\% of the system features.
Most of the features have at least 3 features they depend on and have at least
11 features they influence. The computation of the PageRank~\cite{Brin98} gives
an idea about the ``importance" of single vehicle features and deviates by a factor of almost 20.
This intermeshed structure becomes particularly visible when illustrating the
dependencies as an Edge Bundle View~\cite{Holten06} (see 
\figurename~\ref{fig:edgebundleview}).

\begin{table}[!t]
\renewcommand{\arraystretch}{1.3}
\caption{Distribution of dependencies in the vehicle feature graph}
\label{tbl:distribution}
\centering
%{\scriptsize
\begin{tabular}{|c||p{2cm}|p{2cm}|c|}\hline
&\center{\bf{Dependencies (Ingoing)}}&\center{\bf{Dependencies (Outgoing)}} &
\bf{PageRank}\\
\hline \bf{Maximum} & \center{48} & \center{53}& 5.81\%\\ \hline
\bf{Median}& \center{3} & \center{11} & 0.72\%\\
\hline 
\bf{Minimum} & \center{0} &\center{0}& 0.28\%\\ 
\hline
\end{tabular}
%}
\end{table}

\subsection{Awareness of Dependencies (RQ 2)}
Table~\ref{tbl:awareness} summarizes the results of the expert interviews that
we conducted in order to assess the plausibility and awareness of the analyzed
feature dependencies. 

\begin{table}[!t]
\renewcommand{\arraystretch}{1.3}
\caption{Plausibility and awareness of the analyzed feature dependencies
(n=100)}
\label{tbl:awareness}
\centering
%{\scriptsize
\begin{tabular}{|c||c|c||c|}\hline
&\bf{known}&\bf{unknown} & \bf{sum} \\
\hline \hline
\bf{plausible} & 41.0\% & 48.0\%&89.0\%\\ 
\hline
\bf{implausible}& 1.0\% & 10.0\% &11.0\%\\
\hline\hline
\bf{sum}& 42.0\% & 58.0\% &100\%\\
\hline
\end{tabular}
%}
\vspace*{\dist}
\end{table}

The results indicate that our analysis produced reasonable results as only
11\% of the examined feature dependencies were considered as
\emph{implausible}, i.e., the dependencies were a result of our analysis but the
experts considered them as not correct or at least they were not able to give
account of them. Of the 100 feature dependencies that we examined, 42\% were
known to the experts and 58\% were unknown. Most of the feature dependencies
that we examined were considered as unknown but plausible, i.e., the experts
were not aware of the dependency between the features but when examining the
affected signals and leaf functions they found reasonable explanations for them.
One examined dependency was considered as known and implausible as the expert
were aware of it but had no explanation why this dependency exists.

\subsection{Importance of Dependencies (RQ 3)}
Our interviews reveal that the knowledge about feature dependencies is
especially important for impact analyses on features and signals. The experts
for example mentioned: ``Feature dependencies are important for the assessment
of the complexity, especially when considering the impact of errors'' and ``It
is important to know who uses the signals that features in my responsibility
provide''. However, the interviews also revealed several problems in the
elicitation and revelation of these dependencies. Two main reasons for that are
incomplete documentation and dependencies that arise from architectural
decisions. The experts said: ``Many dependencies arise from specific local
signals that are provided by a central leaf function and used by a lot of
features'' and ``Dependencies between features that are known to function
correctly together are not explicitly documented''. As a major potential
benefit of a rigorous documentation of feature dependencies, the experts named
the precise tracing of logical signals and architectural decision to
requirements. They said: ``Tracing links between requirements and architectural
decisions would be very useful'' and ``A back-link from logical signals to
the requirements that caused them would be beneficial''.

\section{Discussion}
% Before we discuss the results and relate
% them to existing work and evidence, we discuss the threats to validity. 

\subsection{Threats to Validity}
A threat to the internal validity is the fact that the analyzed model is
already a realization/implementation of the system features.
Dependencies might thus be a consequence of a design decision made
by a developer and not an integral part of the system features
itself.
Another threat pertains to the definition of dependency as given in this
paper. Besides the explicitly modeled dependencies that are in the focus of this
paper, there may also be dependencies between vehicle features that occur when
features are implicitly connected through a feedback loop through the
environment.

A threat to the external validity is that we performed this study in a
development and tooling context specific to the \emph{Driving Dynamics and
Driver Assistance} department of the BMW group. This context might not be
transferable to other companies or domains. However, from our experience, we are
confident that the definition of system features that are implemented by a
network of functional blocks is pretty much standard in the development of automotive
software systems.

\subsection{Impact / Implications}
The conclusions we draw point at a number of problems that occur in today's
development of automotive software systems. Current development processes handle
vehicle features more or less as isolated units of functionality~\cite{Broy06}.
This has to some extent historical reasons as the automotive industry managed to make their
different functionality as independent as possible such that vehicles could be
developed and produced in a highly modular way. With the coming up of
software-based functions in the vehicle this independence disappeared~\cite{Broy06}.

Furthermore, our results show that developers consider the knowledge about functional
dependencies as important, especially for tracing purposes and impact analyses. Architectural
decisions hide and scatter these dependencies, which leads to the large number
of unknown feature dependencies as reported in the last section.
%CHANGE
An interesting point is that the reasons for feature dependencies
that were considered as implausible can also be related to architectural
concerns. Leaf functions are architectural elements, which are
subject to reuse and thus related to a number of features. Developers use
leaf functions without considering other vehicle features that might also affect
or be affected by this leaf function. The emerging feature dependencies
were, in most cases, considered as implausible.
%CHANGE END

Therefore, we argue that these dependencies need to be modeled precisely on the
level of vehicle features, still independent from any architectural
design decisions (cf.~\cite{Broy10}).

\subsection{Relation to Existing Evidence}
The results of RQ~1 reflect the results
of a study we have performed with another automotive company, in which we
analyzed the software architecture of a truck~\cite{Vogelsang12}. In that study,
vehicle features showed a comparable extent of dependencies, i.e., at least 69\%
of the analyzed vehicle features depend on other vehicle features or influence other vehicle
features. The analyzed software system of that study was smaller and contained
only 55 vehicle features. 

Our results back up the challenges mentioned
in~\cite{Pretschner07} and \cite{Broy06}, where the authors state that 
%functions {[}of a vehicle{]}
features do not stand alone, but exhibit a high dependency
on each other, so that a vehicle becomes a complex system where all functions act together.

\balance
\section{Conclusions and Future Work}
The results of this paper show that dependencies between vehicle features pose a
great challenge for the development of automotive software systems. Not only
that almost every vehicle feature depends on and/or influences another vehicle feature, we
have also seen that modeling the dependencies on an architectural level is
insufficient for analyzing them, leading to a 50\%
chance that a developer is not aware of a specific dependency. In our study this was
particularly striking when the feature dependencies arose from architectural
decisions.
Considering these conclusions we plan to further discuss our results with the
developers in order to integrate the modeling of dependencies on the
level of vehicle features. Therefore, we have to specify features more
precisely, for example by annotating them with inputs and outputs, and define
the dependencies based on this notion of a vehicle feature (cf.~\cite{Broy10}). We especially plan to integrate features
into a feature hierarchy and describe the dependencies between features by means of a
mode concept~\cite{SPESBook}. This structured specification models dependencies
independently from architectural decisions and thus facilitates the modeling of
feature interactions in requirements engineering.

\bibliographystyle{IEEEtran}
% argument is your BibTeX string definitions and bibliography database(s)
\bibliography{references, IEEEabrv}

% that's all folks
\end{document}